\documentclass[journal]{IEEEtran}

\usepackage{cite}
\usepackage{graphicx}
\graphicspath{{Images/}}
\usepackage{amsmath}
\usepackage{listings}
\usepackage{xcolor}
\usepackage{subfigure}

\usepackage{algorithmic}
\usepackage{multirow}
\usepackage[linesnumbered,ruled,vlined]{algorithm2e}
\usepackage{booktabs}

\newcommand\ohm{\ensuremath\Omega}

\begin{document}


\title{DNA Pattern Matching Acceleration with Analog Resistive CAM}

\author{Jinane Bazzi, Jana Sweidan, Mohammed E. Fouda, Rouwaida Kanj, and Ahmed M. Eltawil
\thanks{J. Bazzi and A. Eltawil are with King Abdullah University of Science and Technology (KAUST), Thuwal 23955, Saudi Arabia.}
\thanks{J. Sweidan and R. Kanj are with ECE Dept., American University of Beirut, Lebanon, 1107 202}
\thanks{M. Fouda is with the Center for Embedded \& Cyber-physical Systems, University of California-Irvine, Irvine, CA, USA 92697-2625. Email: foudam@uci.edu}
\thanks{Manuscript received xxx, xxx; revised xxxx, xxx.}}

\markboth{
}%
{Bazzi \MakeLowercase{\textit{et al.}}: DNA Pattern Matching with aCAM}

\maketitle


\begin{abstract}
DNA pattern matching is essential for many widely used bioinformatics applications. Disease diagnosis is one of these applications, since analyzing changes in DNA sequences can increase our understanding of possible genetic diseases. The remarkable growth in the size of DNA datasets has resulted in challenges in discovering DNA patterns efficiently in terms of run time and power consumption. In this paper, we propose an efficient hardware and software codesign that determines the chance of the occurrence of repeat-expansion diseases using DNA pattern matching. The proposed design parallelizes the DNA pattern matching task using associative memory realized with analog content-addressable memory and implements an algorithm that returns the maximum number of consecutive occurrences of a specific pattern within a DNA sequence. We fully implement all the required hardware circuits with PTM 45-nm technology, and we evaluate the proposed architecture on a practical human DNA dataset. The results show that our design is energy-efficient and significantly accelerates the DNA pattern matching task compared to previous approaches described in the literature.  
\end{abstract}

\begin {IEEEkeywords}
DNA sequencing, Disease diagnosis, Pattern matching, Associative memory, Analog CAM (aCAM).
\end{IEEEkeywords}

\section{Introduction}\label{sec:1}
Deoxyribonucleic acid (DNA) pattern matching is the workhorse for several bioinformatics applications. Disease diagnosis is the most popular among them \cite{mane2016disease}. Scientists rely heavily on DNA pattern matching to explore and detect possible diseases that can arise due to changes in DNA sequences.

A DNA molecule contains the information needed for the development and functioning of organisms. DNA has four basic nucleotide characters: adenine (A), cytosine (C), guanine (G), and thymine (T). A combination of these characters forms a DNA sequence that is unique to each organism. Genes are subsequences of DNA that carry information about an organism's physical traits. It is important to understand and analyze gene sequences since changes in these sequences, referred to as mutations, may have harmful effects on the organism in which they occur, for example, by causing a genetic disorder. One of the major changes is nucleotide repeats, in which a specific DNA pattern repeat is expanded abnormally within a region of DNA. More than 40 repeat-expansion diseases are known, most of which primarily affect the nervous system. Expanded trinucleotide-repeat disorders, which are caused by the unstable expansion of three nucleotides consecutively repeated, were the first to be discovered, and they remain the most frequent. Recently, the list of repeat-expansion diseases has increased to include tetra-, \mbox{penta-,} hexa-, and even dodeca-nucleotide repeat expansions \cite{paulson2018repeat}. Table~\ref{tab:diseases} shows some examples of diseases and their corresponding genes and pattern counts. DNA pattern matching can be used to detect these diseases by identifying the number of consecutive occurrences of the corresponding pattern.

\begin{table*}[tb!]
\centering
\caption{Examples of nucleotide-repeat disorders \cite{diseases1, diseases2}.}
\begin{tabular}{lcccc}
\toprule
\multirow{2}{*}{\textbf{Disease}}  & \multirow{2}{*}{\textbf{Gene}} & \multirow{2}{*}{\textbf{Pattern}} & \textbf{Normal} & \textbf{Disease} \\ 
&&& \textbf{range} & \textbf{range} \\
\midrule
Ataxia syndrome                      & FMR1          & CGG              & 6--54                  & 55--200                 \\ 
Friedreich's ataxia                  & FXN           & GAA              & 5--33                  & 66--1300                \\ 
Huntington's disease                 & HTT           & CAG              & $\leq$26        & $>$40       \\ 
Fragile XE syndrome                  & AFF2          & CCG              & 6--25                  & $>$200      \\ 
Myotonic dystrophy 2                 & DMPK          & CCTG             & 11--26                 & 75--11000              \\ 
Spinocerebellar ataxia 1             & ATXN1         & CAG              & 6--35                  & $\geq$39      \\ 
Huntington's disease-like 2            & JPH3          & CTG              & 6--28                  & 4--60                   \\ 
Spinal and bulbar muscular atrophy   & AR            & CAG              & 11--24                 & 40--62                  \\ 
Dentatorubral-pallidoluysian atrophy & ATN1          & CAG              & 7--25                  & 49--88                  \\ 
Oculopharyngeal muscular dystrophy   & PABPN1        & GCG              & $\leq$10        & 12--17                  \\ 
\bottomrule
\end{tabular}
\label{tab:diseases}
\end{table*}

Datasets of DNA sequences require a huge amount of storage. For instance, the human genome has around 3.1647 billion DNA base pairs \cite{adjeroh2002dna}. This significant volume of DNA data imposes a challenge in performing DNA pattern matching efficiently. Many high-speed DNA pattern matching algorithms have been proposed in the literature. For example, to accelerate search operations with large DNA sequences, in \cite{algorithms}, the authors proposed three efficient algorithms: first-last, processor-aware, and least frequency pattern matching. Although these algorithms showed an improvement in terms of time cost compared to existing algorithms (brute force \cite{BF}, Boyer--Moore \cite{boyer1977fast}, and divide and conquer pattern matching \cite{raju2018parallel}), they still need to be further accelerated by hardware. Other researchers have developed acceleration systems using parallel computing platforms, for instance GPU \cite{gpu} and FPGA \cite{fpga}, to accelerate DNA pattern matching. However, they do not provide a significant improvement, as the existing algorithms utilize slow sequential data processing. As such, there is a need not only to accelerate DNA pattern matching algorithms using hardware accelerators but also to develop efficient hardware-friendly algorithms for this task.

Associative memory (AM) is a powerful tool for in-memory computing. It is a form of storage device that can be searched in a parallel manner \cite{arsovski2003ternary}. The address of any content that matches an input data word is returned. The fast parallel search offered by AM means that this type of memory is used in a variety of applications with big data workloads, for instance, genomic analysis, for which the amount of data has experienced exponential growth in recent years \cite{garzon2022hamming}. 
Content-addressable memory (CAMs) is one way to implement AMs. CAMs can be realized with different technologies, for instance, digital CAMs are implemented using standard CMOS and flip-flops \cite{kokosinski2002fpga}. However, these types of architecture have high power consumption and low density. Accordingly, various emerging resistive device technologies have been used to implement some recently proposed CAMs, benefiting from their non-volatility and high packing density \cite{fouda2022memory}. Examples of resistive devices include ReRAM, phase change memory, magnetic tunneling junctions, and ferroelectric devices \cite{yin2020fecam}. Resistive-based CAMs have an order of magnitude improvement in power consumption and area. Moreover, they can store wide intervals, thereby enabling continuous search ranges for analog applications. In this work, we study memristor-based analog CAM (aCAM), which has a higher memory density and lower power intake. aCAM could accelerate existing applications and may enable potential new uses \cite{nature}.

In this paper, we propose a hardware architecture that can perform DNA pattern matching efficiently with low cost in terms of latency and power consumption. The proposed hardware accelerator uses AM and implements a hardware-friendly version of the processor-aware pattern matching (PAPM) algorithm \cite{algorithms}, in which input words are compared to the entire pattern simultaneously using AM. Our objective is to detect the presence of possible trinucleotide repeat-expansion diseases. The contributions of this paper are summarized in the following points:

\begin{itemize}
    \item We propose a hardware-friendly version of the PAPM algorithm \cite{algorithms} so that DNA pattern matching task can be executed efficiently on a hardware accelerator. The algorithm returns the maximum number of consecutive repeats of a specific length-3 pattern.
    
    \item We propose the full architecture for a hardware accelerator. It uses AM to enable fast parallel searches and matching. We design all the circuits needed to implement the proposed algorithm. 
    
    \item We evaluate the performance of the proposed design by testing it with human DNA sequences and reporting the overall energy and latency. 
\end{itemize}

The remainder of the paper is organized as follows. Section~\ref{sec:2} discusses DNA pattern matching, the PAPM algorithm proposed in \cite{algorithms}, and our proposed hardware-friendly matching algorithm. Section~\ref{sec:3} describes the proposed hardware architecture and its implementation. Section~\ref{sec:4} presents the experimental setup. Section~\ref{sec:5} reports the simulation results in terms of latency and energy. Finally, Section~\ref{sec:6} concludes the work.

\section{DNA Pattern Search}\label{sec:2}
Pattern matching is the process of finding all occurrences of a pattern in a text. In DNA pattern matching, a DNA sequence is scanned to detect the instances of a pattern of nucleotides within it.

\subsection{Related Work}
In \cite{algorithms}, the authors proposed three efficient algorithms to accelerate searches of large DNA sequences. The PAPM algorithm is a word-based algorithm, in which two words consisting of several characters are compared concurrently. The word length (\texttt{word\_len}) depends on the number of bits per register in the processor. For instance, a 32-bit processor can simultaneously compare length-4 words.

In this algorithm, the text to be searched is a DNA sequence of length $n$, which is stored in an array $t[0 \dots n-1]$. The search pattern has length $m$ and is stored in an array $p[0 \dots m-1]$.
The operation of this algorithm has two phases:
\begin{itemize}
    \item Preprocessing phase: The algorithm scans the interval $t[0 \dots n-m]$ of the text from left to right, searching for the first word in the pattern $p[0 \dots \texttt{word\_len}-1]$. In each iteration, it moves forward one character. When an instance is found, the window of $m$ characters starting with this word is recognized as a potential interval of the text that needs to be matched with the pattern in the next phase.
    
    \item Matching phase: The remaining words in the identified windows are precisely compared with those of the pattern to detect the matches.

\end{itemize}

\subsection{Proposed Hardware-friendly Matching Algorithm}
In our work, we propose a hardware accelerator that uses AM to analyze DNA data more efficiently in parallel. We also propose a modified version of the PAPM algorithm \cite{algorithms} that is executed by this accelerator. The proposed algorithm returns the maximum number of consecutive pattern repeats instead of the total number of occurrences, because our objective is to detect possible repeat-expansion diseases where a specific pattern is abnormally expanded.

\begin{figure}[tb]
\centering
\includegraphics[width=0.77\columnwidth]{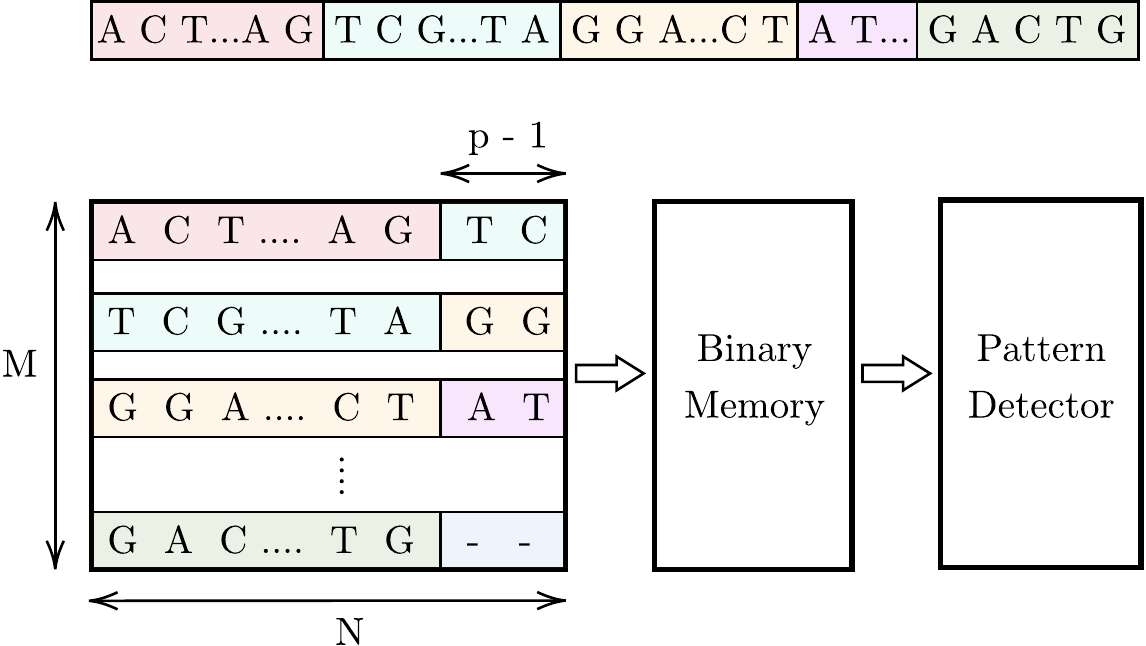}
\caption{Proposed high-level design for comparing a pattern of length 3.}
\label{fig:dna}
\vspace{-0.05in}
\end{figure}

As AM allows parallel searches, we store the DNA data in an AM array, so that in each cycle, we compare a pattern of length $p$ to $p$ cells simultaneously across all rows. For the pattern occurrences found, if any, we store a 1 in the corresponding binary memory cells. Then, we move one character (one cell) and re-compare. As the pattern may be split between two rows, at the end of each row, we use ($p-1$) additional cells to store the first ($p-1$) characters of the row below it, as illustrated in Fig.~\ref{fig:dna}. For the last row, the additional cells store a dummy character that will be explained later.
We then use a pattern detector to find the maximum number of consecutive pattern instances in the stored DNA data. Algorithm~\ref{alg:1} illustrates the steps in the proposed procedure. The pattern detector algorithm is described in Section~\ref{sec:PD}. 

\begin{algorithm}
\caption{Matching phase}\label{alg:1}
\SetKwInput{KwInput}{Input}                
\SetKwInput{KwOutput}{Output}              

\KwInput{DNA data text[1...t] and pattern P[1...p]}
\KwOutput{Memory array mem(M, N-(p-1))}

AM = zeros(M, N)

mem = zeros(M, N-(p-1))

\tcc{Loading Data} 

\tcc{Except last row} 
count=1

\For{$i \gets 1$  \KwTo $N-1$} {
    AM(i, 1:N-(p-1)) =  text(count:count+N-p) \\
    count = count+N-(p-1)
}

\tcc{Last row} 

\tcc{Remaining characters} 
r = t - (M-1)*(N-(p-1))

AM(M, 1:r) = text((M-1)*(N-(p-1)+1:t))

\tcc{Replication of characters} 

AM[(1:M-1), (N+1:N+p-1)] = AM[(2:M), (1:p-1)]

\tcc{N-(p-1) cycles}   

\For{$i \gets 1$  \KwTo $N-(p-1)$} {%

    mem[(1:M), i] = (AM[(1:M), (i:i+p-1)] == P)}
 
\tcc{Run pattern detector}
Pattern\_Detector(mem)
\end{algorithm}

\section{Hardware Architecture}\label{sec:3}
To perform binary and ternary search operations more efficiently, researchers have  recently suggested replacing the CMOS devices in AM with memristors. More recently, the authors in \cite{nature} proposed a form of memristor-based aCAM that can store a range of values, which can be matched or not to an analog input. In addition to its ability to store wide continuous intervals, an aCAM can store multiple narrow ranges as discrete levels, allowing the storage of at least 3 bits per cell. aCAM can replace digital CAM while improving both storage density and power consumption. As such, aCAM is a strong candidate for DNA pattern matching since its ability to store discrete multi-bit ranges enables the search for DNA characters.

\subsection{aCAM Cell}
\label{aCAM}
\begin{figure} [!b]
\centering
\subfigure[]{\includegraphics[width=0.87\linewidth]{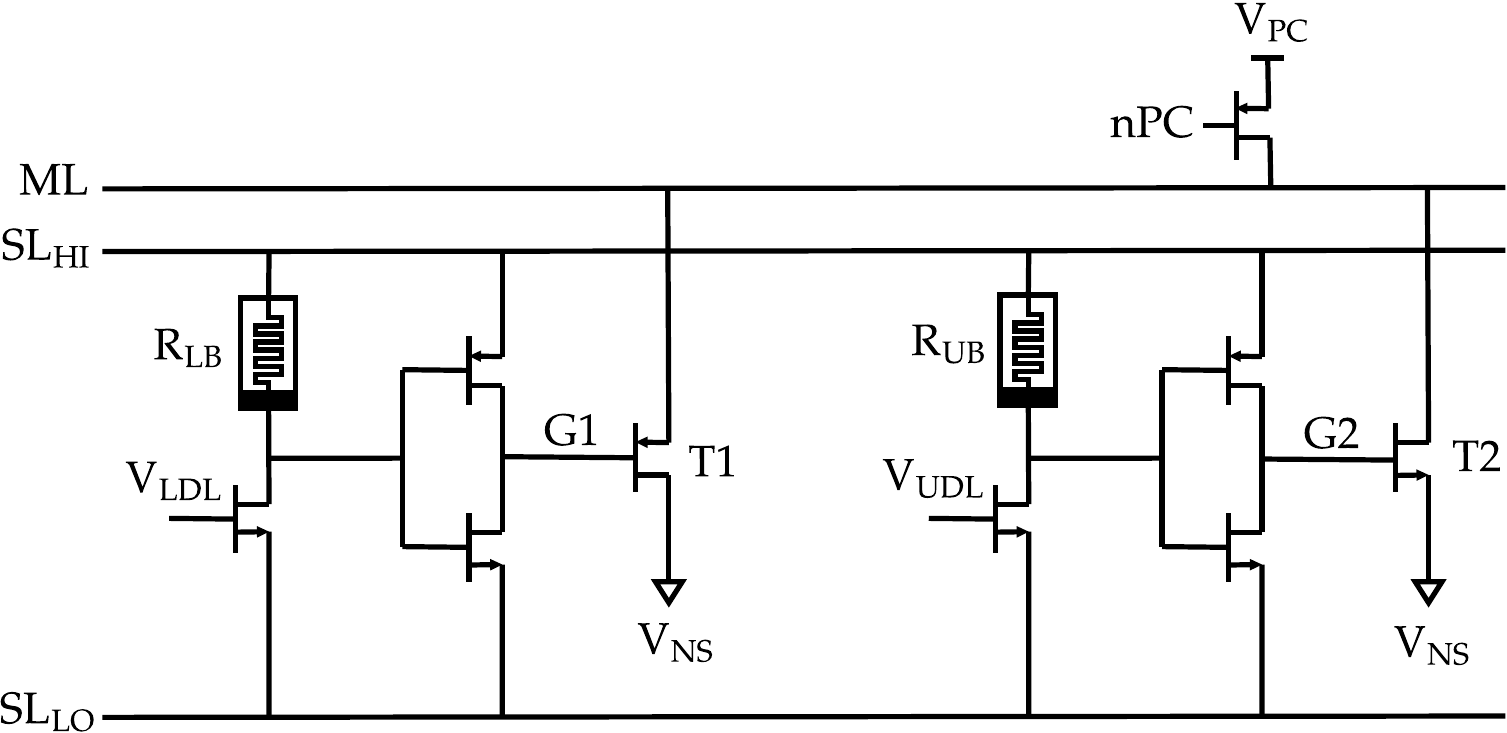}} 
\subfigure[]{\includegraphics[width=1\linewidth]{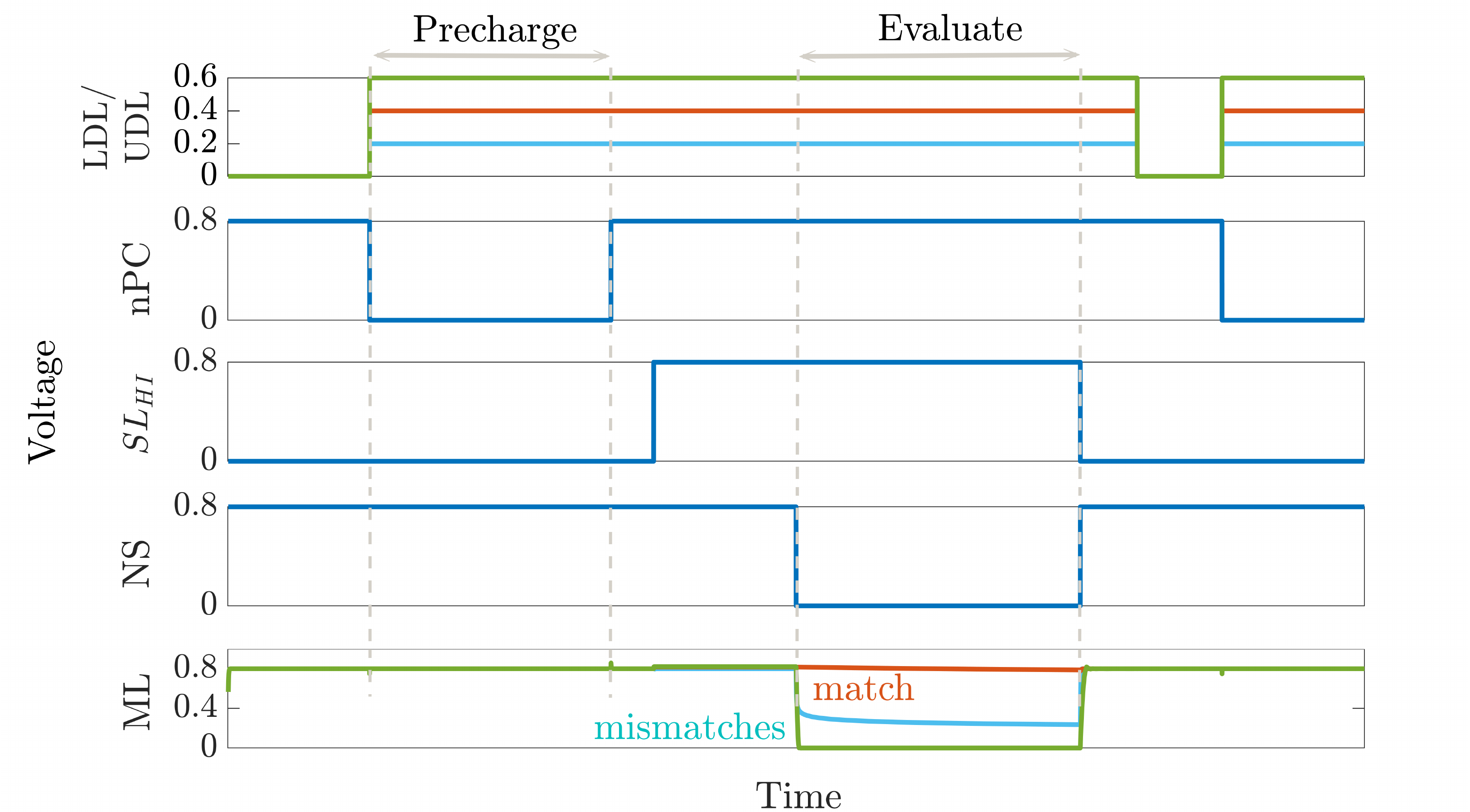}}
\caption{(a) Schematic of a complete 8T2M aCAM cell. (b) Plot of the transient behavior of an 8T2M aCAM cell with match interval [0.35~V, 0.45~V] at three different LDL = UDL inputs: matching, lower-bound mismatching, and upper-bound mismatching.}
\label{fig:aCAM}
\end{figure}

A recent study \cite{materials} compared different types of switching devices including memristors, phase change, magnetorsistors, and FeFET devices. Memristor devices have key properties, such as a high number of distinguishable states, fast switching speed, and good endurance and retention, thereby qualifying them as good candidates for aCAM applications. Several types of memristor-based aCAM have been proposed in the literature. In \cite{bazzi2022efficient}, the authors analyzed and compared different memristor-based aCAM cells. They found that the 8T2M cell was the most energy-efficient and offers low latency, low failure probability, and small area compared to other types of aCAM. Thus, the 8T2M design is a suitable candidate for DNA pattern matching.

An 8T2M aCAM cell is shown in Fig.~\ref{fig:aCAM}(a). The circuit is mainly divided into two voltage divider subcircuits that determine the lower and upper bounds of the stored interval, respectively, depending on the programmed resistance value in each of them. The lower-bound subcircuit consists of a transistor connected in series to a variable resistance $R_{LB}$, followed by an inverter that controls the voltage $V_{G1}$ at the gate of a PMOS transistor T1. The lower-bound match threshold of the aCAM cell (LB) is determined by tuning the resistance $R_{LB}$. Similarly, the upper-bound match threshold (UB) is configured with an independent voltage divider using a variable resistance $R_{UB}$ and an inverter to generate the voltage $V_{G2}$ at the gate of the pull-down NMOS transistor T2. We modified the aCAM cell such that the input for the lower-bound subcircuit $V_{LDL}$ is different from that of the upper-bound subcircuit $V_{UDL}$, for reasons that we explain later.

An aCAM search operation consists of two phases: pre-charge and evaluate. First, the match line (ML) is pre-charged to a high logic level trough a pull-up PMOS transistor by setting its gate voltage nPC to low, as illustrated in Fig.~\ref{fig:aCAM}(b). Then, the evaluation phase starts by setting the $V_{SL_{HI}}$ signal to high and $V_{NS}$ to 0. The arrival of the low NS signal is delayed to give time for G1 to evaluate and prevent it from falsely discharging the ML.

If $V_{LDL}$ is less than the lower bound, a low voltage, smaller than $V_{PC}-|V_{tp}|$, builds on G1, where $V_{PC}$ is the ML pre-charge voltage. This yields a mismatch because T1 will turn on and discharge the ML. On the other hand, if $V_{UDL}$ is greater than the upper bound, a high voltage, greater than the threshold voltage $V_{tn}$, builds on G2, turning T2 on, and the ML is discharged to a low level through T2. Thus, the ML stays high only when the search input belongs to the interval stored in the cell (match), set by the resistance values of $R_{LB}$ and $R_{UB}$.

\begin{figure}[!tb]
    \centering
       \subfigure[]{\includegraphics[width=0.75\columnwidth]{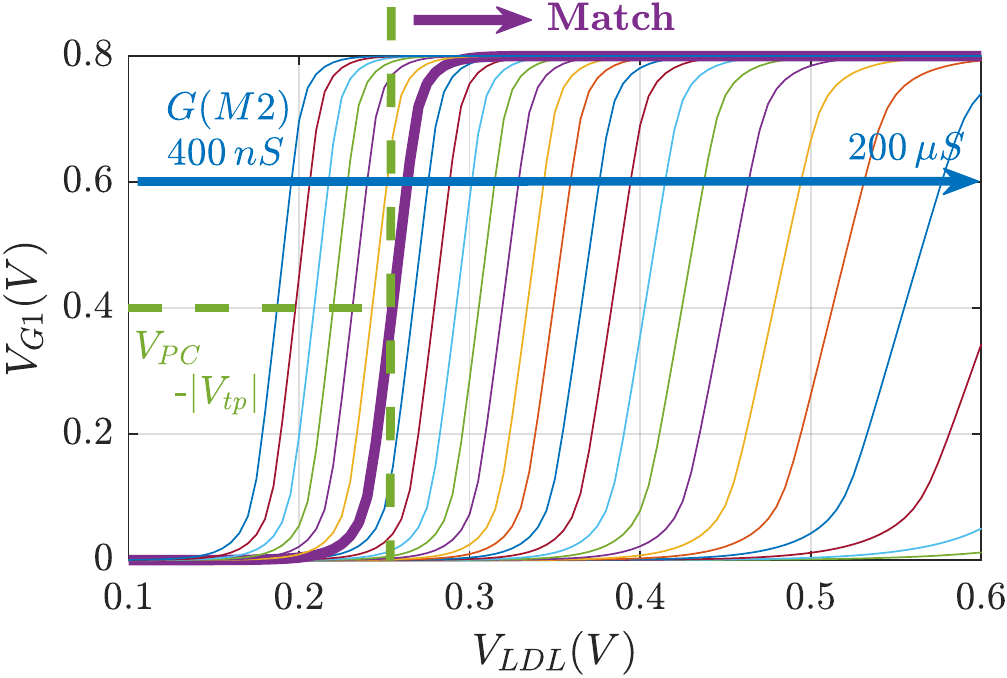}}
       \subfigure[]{\includegraphics[width=0.75\columnwidth]{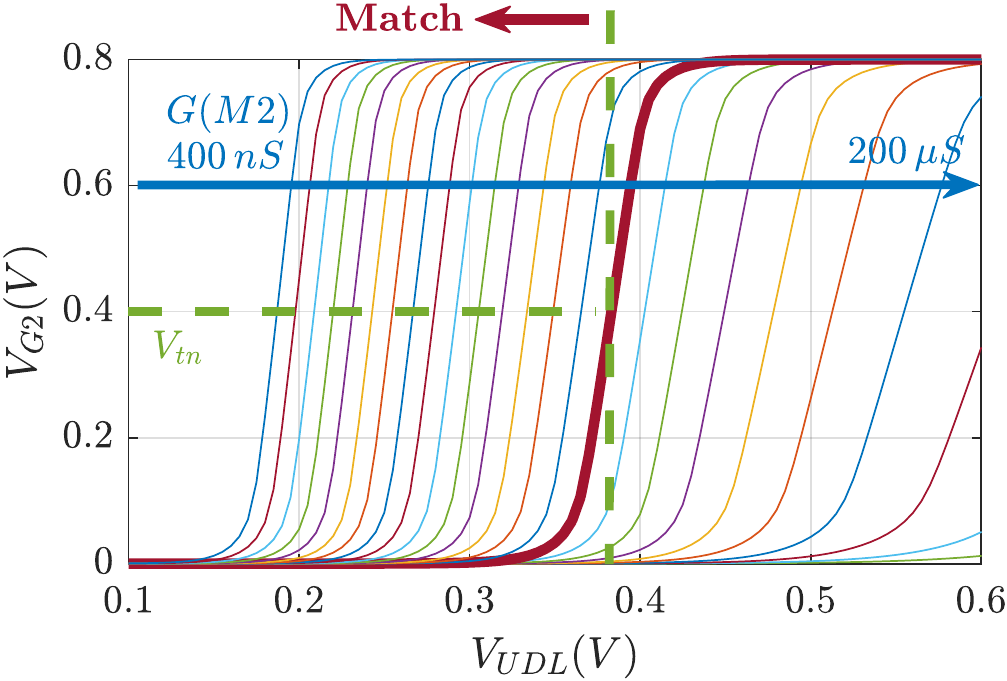}}
    \caption{Voltage transfer characteristics (a) $V_{G1}$ vs $V_{LDL}$ and (b) $V_{G2}$ vs $V_{UDL}$ for $R \in$ [5~k\ohm, 2.5~M\ohm].}
    \label{fig:VG vs VDL}
\end{figure}

Fig.~\ref{fig:VG vs VDL} illustrates the voltage transfer characteristics for $V_G$ versus $V_{DL}$. Based on these curves, for a given pair of $R_{LB}$ and $R_{UB}$ resistances, the corresponding lower and upper bounds of the stored interval can be determined based on the $V_{DL}$ values, which results in $V_{G1}=V_{PC}-|V_{tp}|$ and $V_{G2}=V_{tn}$.

\subsection{DNA Pattern Encoding}

\begin{figure} [!htb]
\centering
\includegraphics[width=0.85\columnwidth]{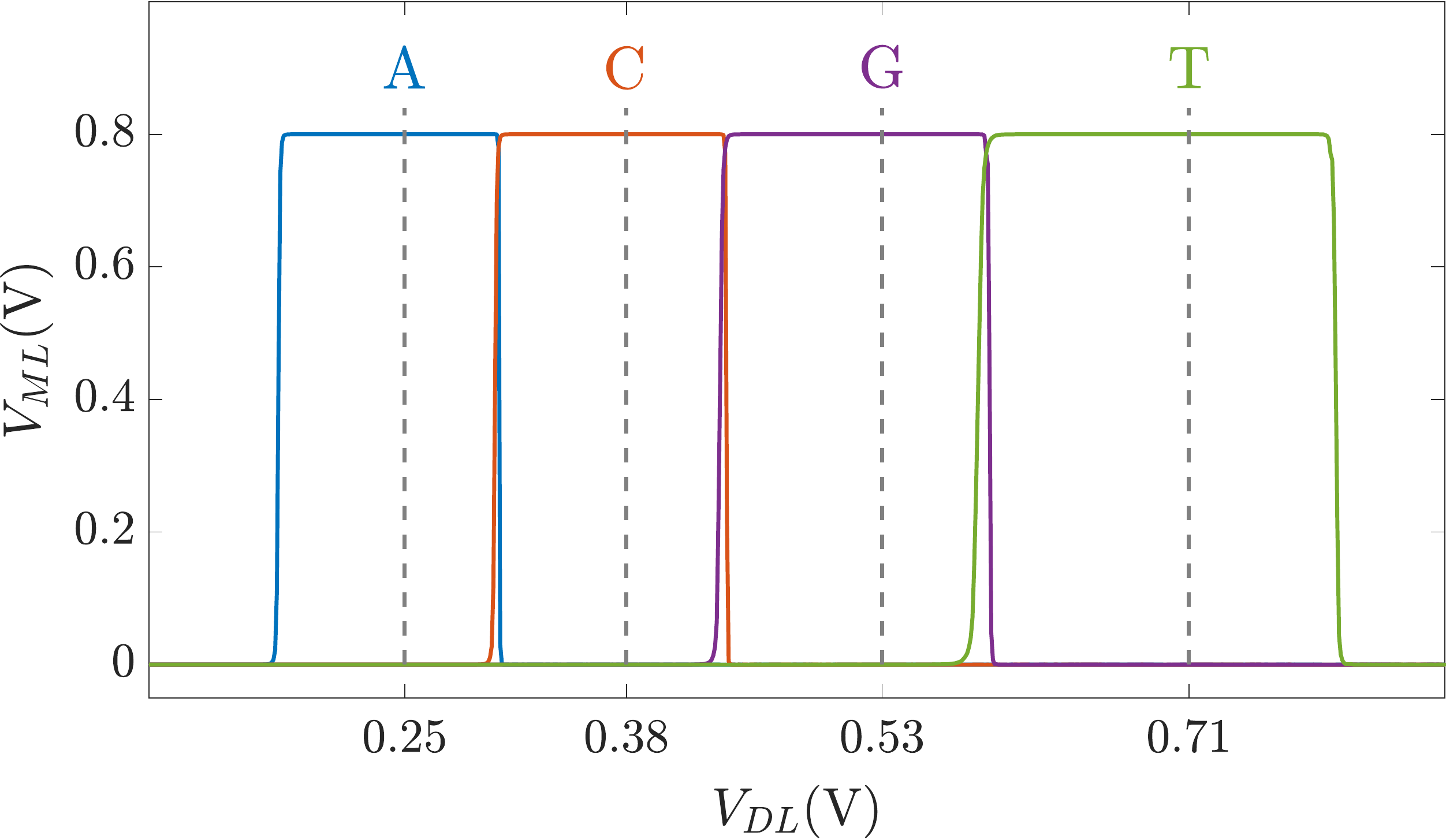}
\caption{$V_{ML}$ measured 
at 1~ns vs $V_{DL}$ for the four DNA character intervals defined in Table~\ref{tab:mapping}.}
\label{fig:vml_vs_vdl}
\end{figure}

As discussed previously, DNA sequences are combinations of four nucleotides: A, C, G, and T. Thus, if one aCAM cell stores one DNA character, four discrete levels are needed to represent the DNA characters. Using the aCAM design shown in Fig.~\ref{fig:aCAM}(a), four intervals are obtained for the resistance configurations shown in Table~\ref{tab:mapping}. Such an encoding guarantees zero failure. Squeezing more than one character per interval would lead to some failure, especially if there was variability \cite{bazzi2022efficient}. Each DNA character can be mapped into one interval defined by a lower and an upper bound, corresponding to resistances $R_{LB}$ and $R_{UB}$, respectively (explained in Section~\ref{aCAM}). Fig.~\ref{fig:vml_vs_vdl} shows the transient ML voltage measured at 1~ns vs $V_{DL}$ for the four intervals. Note that we added a buffer of two inverters after which we read $V_{ML}$, as will be discussed later. The intervals are no longer symmetric because of the gain after the inverters.

\begin{table}[!htb]
\centering
\caption{Nucleotide character encoding}
\begin{tabular}{cccc}
\toprule
Stored                                       & \multicolumn{2}{c}{Programmed resistances} & Stored interval \\ 
character & \multicolumn{1}{c}{$R_{LB}$ (k$\Omega$)}       & {$R_{UB}$ (k$\Omega$)}         & {[LB, UB] (V)}   \\ 
\midrule
A                & 2500        & 186.32      & {[}0.19, 0.31{]}    \\ 
C                & 163.3       & 27.6        & {[}0.32, 0.44{]}    \\ 
G                & 24.9        & 9.69        & {[}0.46, 0.59{]}    \\ 
T                & 8.9         & 5.06        & {[}0.63, 0.79{]}    \\ 
\bottomrule
\end{tabular}
\label{tab:mapping}
\end{table}

\subsection{Row Matching}
To search for a specific character in an aCAM cell, we apply to its inputs a voltage equal to the average of the lower and upper bounds of the corresponding interval defined in Table~\ref{tab:mapping}. The cell matches only if its stored character is equal to the one searched for. 

\begin{table}[tbh!]
\centering
\caption{Different combinations of searched and stored data that result in either a match or mismatch state.}
\resizebox{\columnwidth}{!}
{
\begin{tabular}{ccccc}
\toprule
\multicolumn{3}{c}{Search data}   \\ 
\cmidrule{1-3}
{Mask}               &           & \{$V_{LDL}, V_{UDL}$\} (V)              & Stored character & Output state \\ 
\midrule
{0}                  & --                  & \{$V_{DD}$, 0\}                 & {X}                & {Match}        \\[1ex]
{\multirow{8}{*}{1}} & {\multirow{2}{*}{A}} & \multirow{2}{*}{\{0.25, 0.25\}} & {A}                & {Match}        \\ 
                   &                    &                                 & {C / G / T}        & {Mismatch}     \\[1ex] 
                   & {\multirow{2}{*}{C}} & \multirow{2}{*}{\{0.38, 0.38\}} & {C}                & {Match}        \\ 
                   &                    &                                 & {A / G / T}        & {Mismatch}     \\[1ex]
                   & {\multirow{2}{*}{G}} & \multirow{2}{*}{\{0.53, 0.53\}} & {G}                & {Match}        \\ 
                   &                    &                                 & {A / C / T}        & {Mismatch}     \\[1ex]
                   & {\multirow{2}{*}{T}} & \multirow{2}{*}{\{0.71, 0.71\}} & {T}                & {Match}        \\ 
{}                   & {}                   &                                 & {A / C / G}        & {Mismatch}     \\ 
\bottomrule
\end{tabular}
}
\label{tab:summary}
\end{table}

\begin{figure} [!htb]
\centering
\includegraphics[width=0.9\columnwidth]{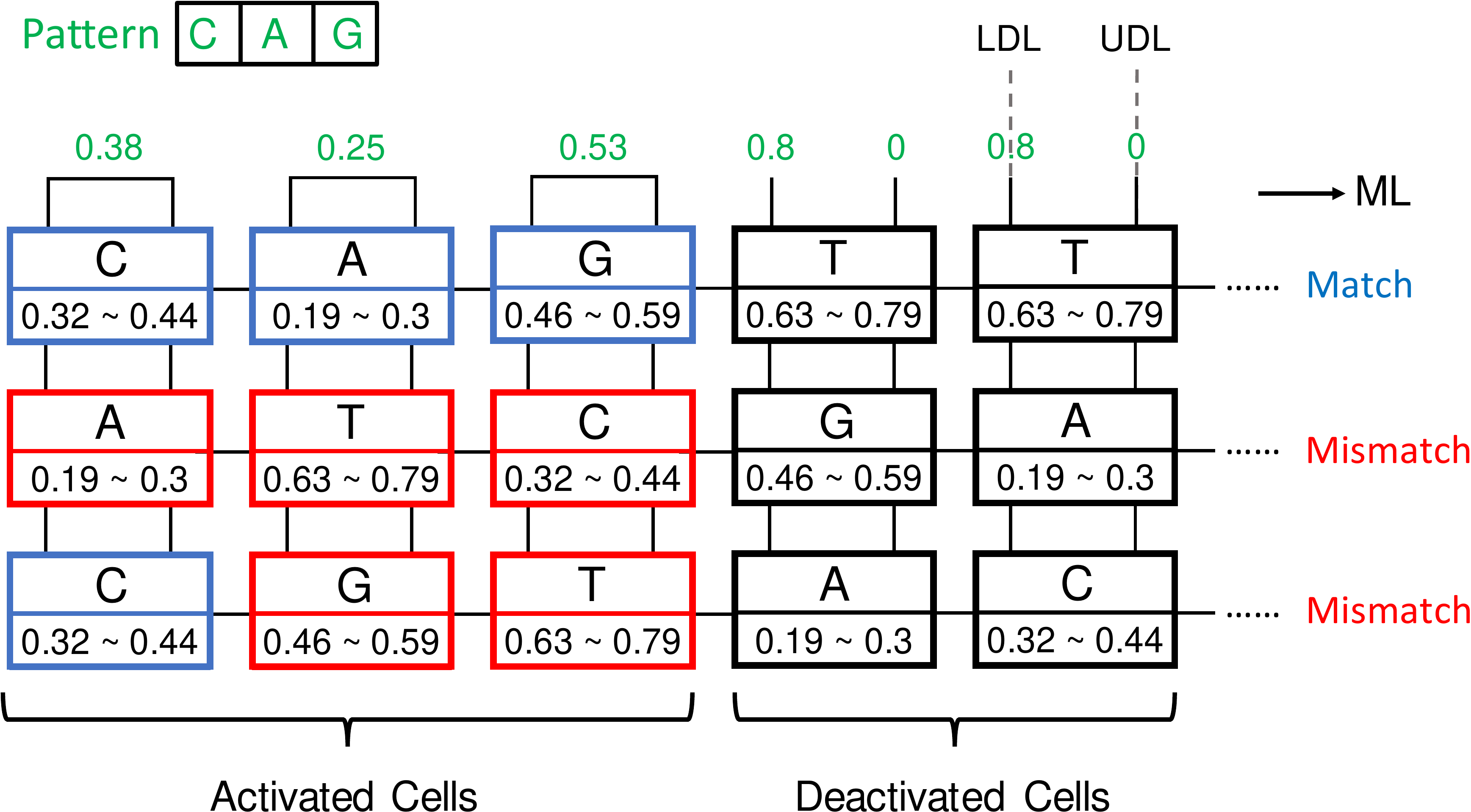}
\caption{aCAM search operation example. Blue cells are matched, red cells are mismatched, and black cells are deactivated.}
\label{fig:example}
\end{figure}

An aCAM cell can be deactivated to give a match, regardless of its stored character, by applying to its lower-bound subcircuit a voltage $V_{LDL}=V_{DD}$ and to its upper-bound subcircuit a voltage $V_{UDL}=0$. Based on Fig.~\ref{fig:VG vs VDL}(a), if the input $V_{LDL}$ is equal to $V_{DD}$ for any resistance value in the range, the resulting $V_{G1}$ is high, greater than $V_{PC}-|V_{tp}|$, thus T1 will remain off and the lower-bound subcircuit matches. Similarly, Fig.~\ref{fig:VG vs VDL}(b) shows that for an input $V_{UDL} = 0$, the obtained $V_{G2}$ is less than $V_{tn}$ for all resistance values. Hence, T2 remains off and the upper-bound subcircuit matches. Table~\ref{tab:summary} shows the different combinations of searched and stored data that result in either a match or mismatch state.
A full match occurs when all the aCAM cells match within the same row, otherwise, a mismatch is recorded, as illustrated in Fig.~\ref{fig:example}.
An additional interval, MM, is needed to give a mismatch for any input. This interval is stored in the extra aCAM cells of the last row of the array, as discussed earlier. We set its $R_{LB}$ to $R_{UB}$ of the last interval and its $R_{UB}$ to $R_{LB}$ of the first interval. 

\begin{figure*}[!tbh]
\centering
\includegraphics[width=1.95\columnwidth]{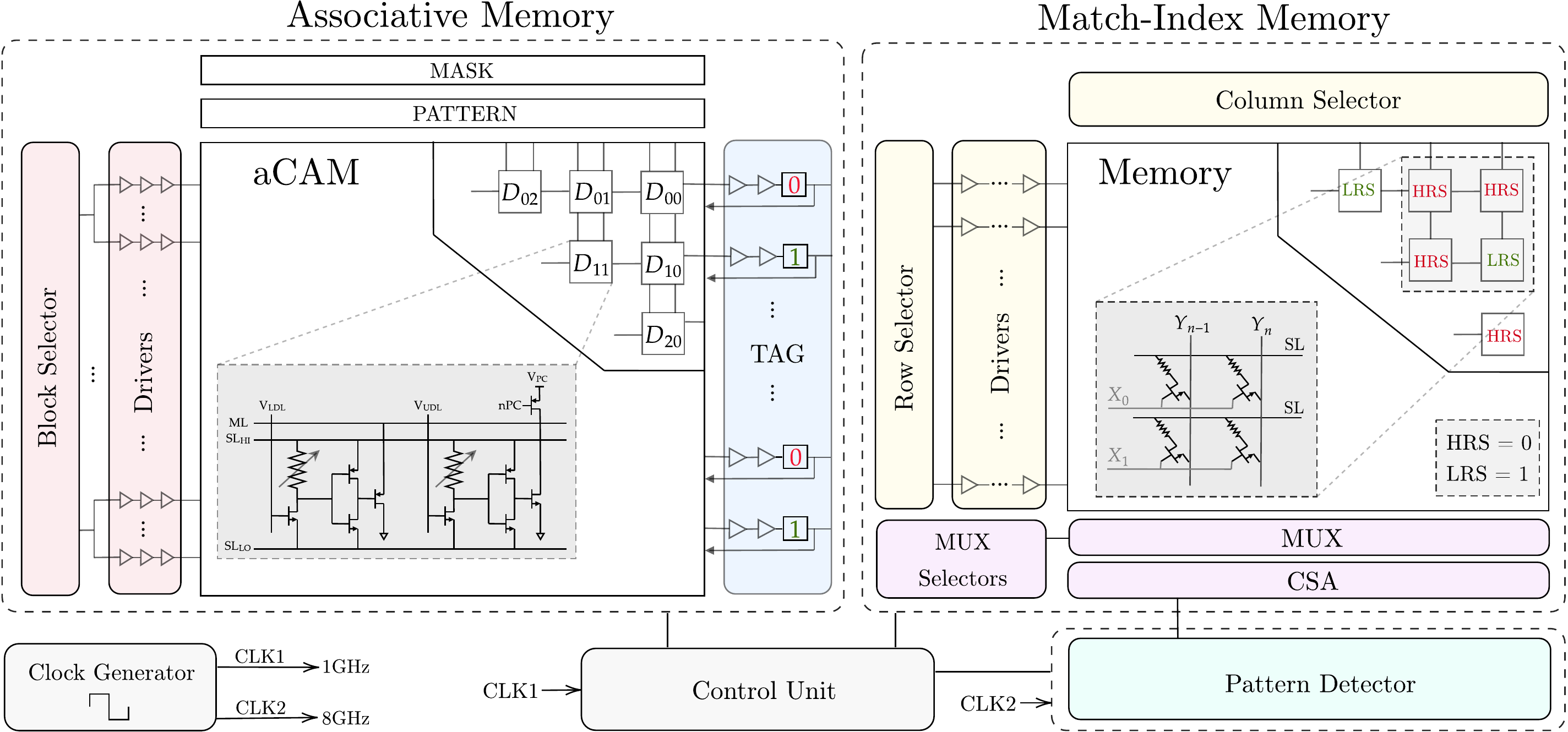}
\caption{Full hardware architecture for the proposed design.}
\label{fig:full_architecture}
\end{figure*} 

\subsection{Proposed Architecture}

Fig.~\ref{fig:full_architecture} is an overview of the proposed architecture. The architecture consists of three main building blocks, as discussed earlier: (1) AM, which stores the DNA sequence, (2) matching-index memory, which stores the index of the pattern matches, and (3) the pattern detector, which finds the maximum number of consecutive pattern repeats. The details of the full implementation of each block are described in this section.

\subsubsection{Associative Memory}
The designed AM consists mainly of three blocks: aCAM array, block selector, and tag registers.

\paragraph{aCAM array}
The DNA data are stored in an $M \times N$ aCAM array. Since we are looking for the occurrences of a specific pattern of length equal to $p$, then, in our search, we activate a window of only $p$ cells to check if this pattern exists. As the comparison can be done in parallel across the rows, $p$ cells are activated per row in each cycle, and these are compared to the pattern. The input to the activated cells is set corresponding to the pattern characters, as defined in Table~\ref{tab:summary}, whereas the aCAM cells of the deactivated columns will have $V_{LDL}$ set to $V_{DD}$ and $V_{UDL}$ set to 0, so that they result in a match, regardless of the characters they are storing, as discussed earlier. The window of $p$ activated cells is shifted by one cell in each cycle till the end of the row, where there are special cases, since the pattern may be split between two rows. To handle these cases, we replicate $(p-1)$ characters. Thus, at the end of each row, we replicate and add the first $(p-1)$ aCAM cells from the row below it. If the DNA data are not a multiple of the row length $N$, the extra cells store the interval MM so that they always result in a mismatch.

\paragraph{Block selector}
In nucleotide-repeat diseases, the location of the pattern repeats is important. In fact, for each disease, the repeats must be in a specific gene. For example, for Huntington's disease, the CAG repeats must be in the HTT gene, which is a portion of the entire DNA sequence. To allow the detection of different diseases, we store multiple critical genes in the aCAM array. We divide the array into eight blocks, so that for each disorder, we activate the blocks containing the corresponding gene. To select or activate specific blocks, we designed a block selector. Its implementation is explained in Supplementary Note~1. The block selector activates the desired blocks by connecting its output to the aCAM cells as follows. Fig.~\ref{fig:aCAM}(b) shows the signaling for the 8T2M aCAM design. In the evaluate phase, $V_{NS}$ is set to 0. An aCAM cell can, therefore, be deactivated by keeping its $V_{NS}$ high during this cycle. As such, the aCAM cells of an activated block should have $V_{NS}=0$ during the evaluation, whereas those included in a deactivated block should always have $V_{NS}$ high. Therefore, the block selector output should be inverted before it is connected to the NS of an aCAM cell.

\paragraph{Drivers}
The output of each block selector is connected to several chains of three inverters with an increasing width size, and their output is connected to the NS of the aCAM cells.

\paragraph{Tag registers}
In each cycle, the ML result of each row is stored in the tag, which is written to memory. The tag consists of a buffer (two inverters) connected to a flip-flop that stores the ML value.

\subsubsection{Match-Index Memory}
In each cycle, after we have the values ready at the tag, we write them to a 1T1R memory array, so that we can later read the memory and input the values to the pattern detector to detect the patterns.
The role of each 1T1R cell is to store either a low-resistive state or a high-resistive state (HRS) corresponding to whether the pattern in the aCAM array is a match or mismatch. This is done during the memory write operation. After the array is populated, the next step is to read the array. The pattern detector counts how many consecutive patterns there are and determines the length of the sequence.
After these operation are complete, we reset the memory array to HRS so that it is for the next use. The control unit generates the read, write, and reset signals to organize the sequential functions of the memory.

\begin{figure*}[tb]
\centering
\includegraphics[width=1.6\columnwidth]{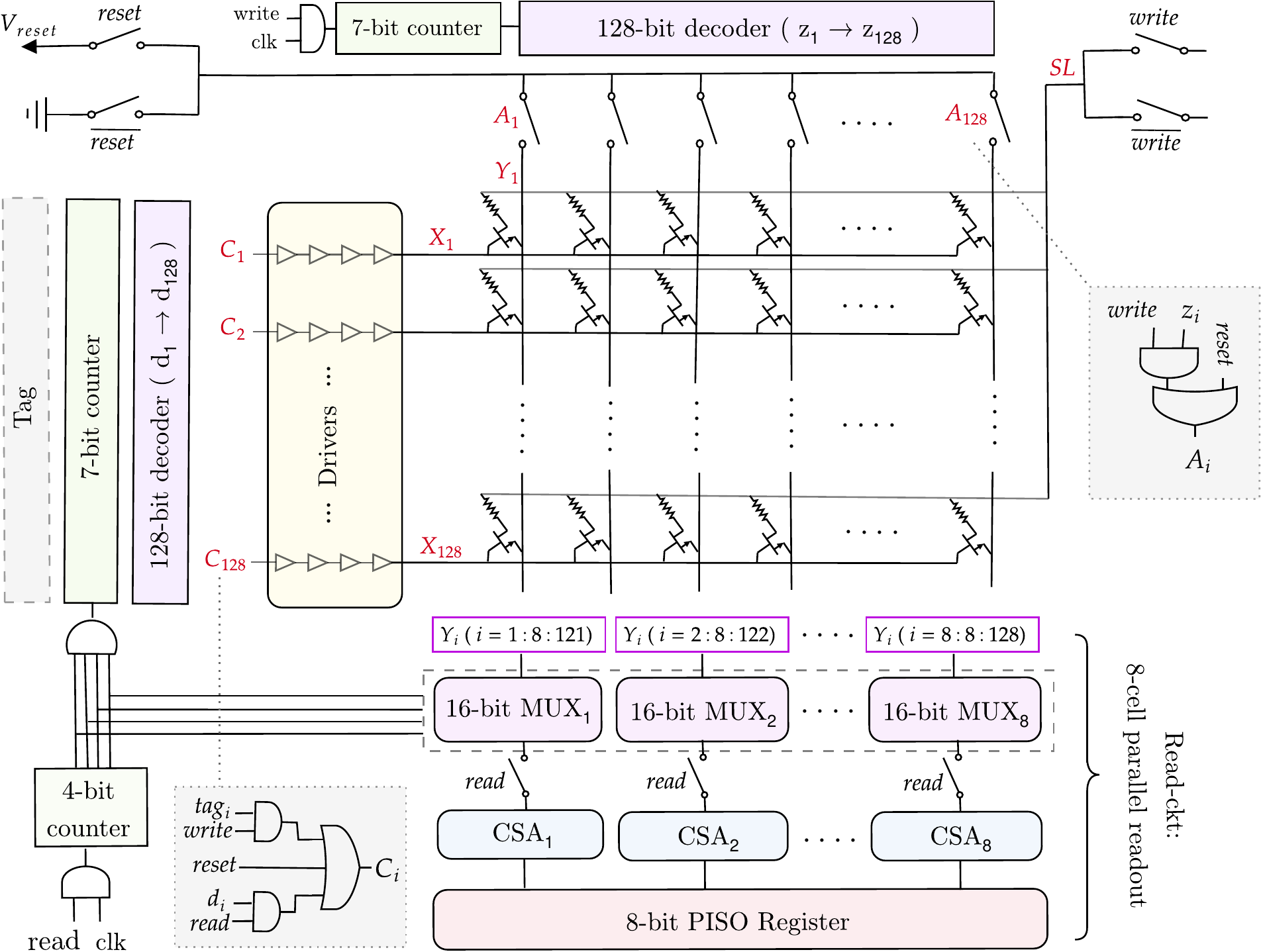}
\caption{Full memory architecture for a $128 \times 128$ match-index array.}
\label{fig:mem_arch}
\end{figure*}

Each 1T1R cell consists of a memristor and an NMOS transistor, connected as shown in Fig.~\ref{fig:mem_arch}. The values of the signals at each node (SL, X, and Y) vary depending on the write, read, or reset mode of the memory, as summarized in Supplementary Table~III.

\paragraph{Memory write}
During the memory write operation, the rows of the 1T1R memory array are filled in parallel based on the values received from the tag in each cycle. Each tag output is connected to a corresponding row (X), which is determined by the signal C, as illustrated in Fig.~\ref{fig:mem_arch}. This is then followed by drivers with four inverters, which drive the current. As all rows are written in parallel, a column selector circuit is needed to write the columns (Y) of the memory array sequentially. 

As shown in Supplementary Table~III, to write a column, the corresponding Y node must be 0 and all the others 1 to avoid creating a current path between Y and SL (since SL = $V_\text{set}$) and thus, to avoid writing them as well. To achieve this functionality, we need a counter followed by a decoder and transmission gates. In this way, the transmission gate outputs a 0 only for the targeted column that is to be written. All the other columns are high. The transmission gates are controlled by the signal A (Fig.~\ref{fig:mem_arch}), which passes the ground value for the specific column chosen by the decoder ($z=1$). Note that while writing the memory, the read circuitry is disconnected using the transmission gates controlled by the read signal.

For an $n$-column memory, the column selector circuit has a $\log_{2}n$-bit counter, then a $\log_{2}n$ to $n$-bit decoder, followed by $n$ transmission gates to get the desired Y value. This architecture is illustrated for a $128 \times 128$ memory in Fig.~\ref{fig:mem_arch}.

%

\paragraph{Memory read}
When all the write cycles are complete, then the 1T1R memory array is fully written. We need to read the values and input them, one by one, to the pattern detector circuit, which counts and detects the maximum sequence length. Reading the whole memory requires a row selector to traverse the rows and a multiplexer to connect the columns to a current sense amplifier sequentially to sense the stored values. 

To avoid traversing the whole memory cell by cell and thus, increasing the delay, we will read every eight consecutive cells in parallel and save them in an 8-bit parallel-in serial-out (PISO) register, which will output them serially, one by one, to the pattern detector circuit. To do so, we need eight $n/8$-bit multiplexers, which are a chain of transmission gates whose function is to select the eight columns and connect them to the corresponding current sense amplifier. Thus, we need eight current sense amplifiers for the eight columns selected. Their output will be passed to the 8-bit PISO register, as shown in Fig.~\ref{fig:mem_arch}. 
The selectors of the multiplexers are determined by a $\log_{2}n/8$-bit counter. The outputs of this counter are AND'ed and connected as the clock to the row selector circuit counter, which ensures that we read the next row only when all the current columns are read. Note that during the read operation, the column selector circuit is disconnected.

For an $m \times n$ memory array, the row selector circuit consists of a $\log_{2}m$-bit counter, which is followed by a $\log_{2}m$- to $m$-bit decoder, then drivers to drive the current and $m$ transmission gates to get the desired X node value.

\paragraph{Memory reset}
Finally, as shown in Fig.~\ref{fig:mem_arch}, to reset the memory after each memory read, the columns (Y) must be connected to $V_\text{reset}$, SL to the ground, and the rows (X) to the reset signal to turn on the NMOS transistors and allow the current to flow from Y to SL, thus resetting the 1T1R cell to a HRS. During this mode, the column and row selectors and the read circuits are all disconnected.

%

\subsubsection{Pattern Detector}\label{sec:PD}
The output of the PISO register is input, bit by bit, into the pattern detector, which finds the maximum number of consecutive instances of a specific pattern. A bit equal to 1 indicates that an occurrence of the pattern was found. Thus, for patterns consisting of different characters, for example CAG, every 1 should be followed by $(p-1)$ zeros. As such, whenever a 1 is received, the pattern detector can skip $(p-1)$ bits and check again. If the input is 1, the number of repeats is incremented. If it is a 0, the maximum is updated, and the count restarts from 0 and the sequence of inputs is traversed, bit by bit, until a new 1 is found, after which we can skip $(p-1)$ bits and check again. Note that for DNA pattern detection, the maximum length of a pattern consisting of different characters is 4, as there are only four nucleotides.

\begin{figure} [!tb]
\centering
\includegraphics[width=0.85\columnwidth]{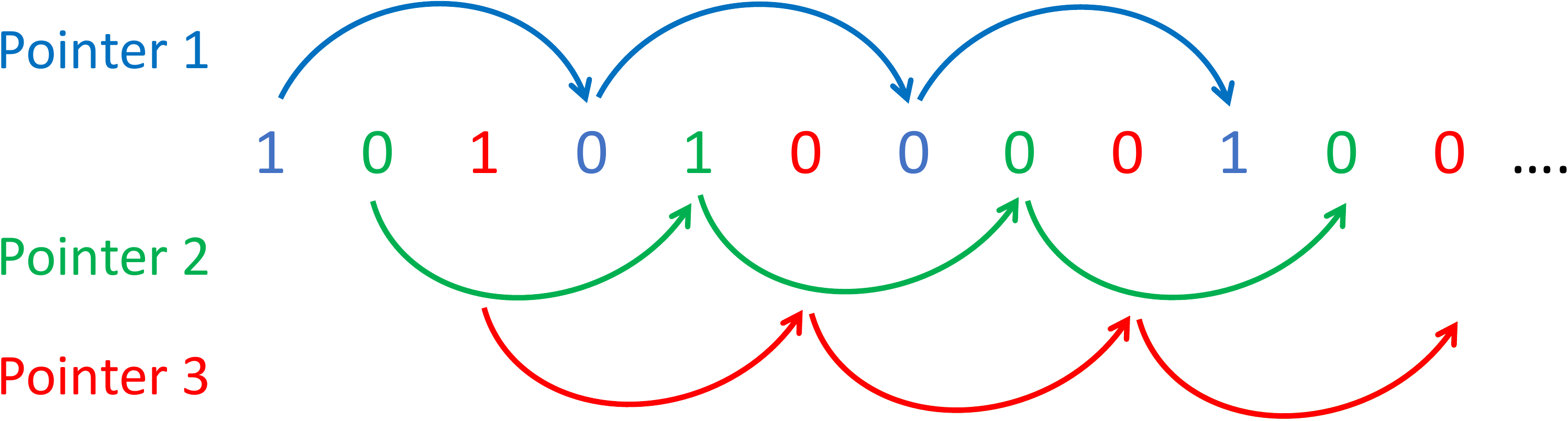}
\caption{Three pointers traversing the sequence of memory bits.}
\label{fig:pointers}
\end{figure}

To generalize the pattern matching procedure to any DNA pattern, more than one pointer must traverse the sequence, because a 1 is not necessarily followed by $(p-1)$ zeros. Without loss of generality, for length-3 patterns, three pointers are needed to scan the inputs received from the PISO register. The first one checks the inputs having an index modulo $3=1$. The second pointer scans the inputs with index modulo $3=2$, and finally, the third pointer traverses the inputs with index modulo $3=0$, as illustrated in Fig.~\ref{fig:pointers}. Pointer 1 starts from the first input, then it skips 2 bits to check the fourth input, and so on, and similarly for pointers 2 and 3. Each pointer has its own counter and maximum register, which saves the number of consecutive repeats. 

\begin{figure}[!tb]
\centering
\includegraphics[width=0.9\columnwidth]{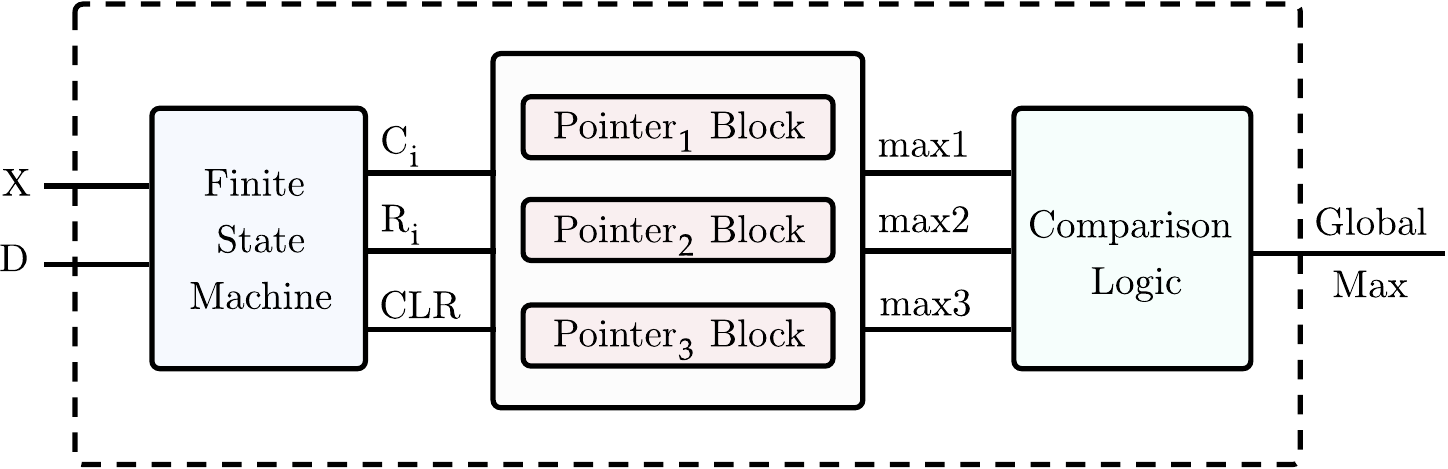}
\caption{Block diagram of a pattern detector for comparing a length-3 pattern.}
\label{fig:pattern_detector}
\end{figure}

\begin{figure}[!tb]
\centering
\includegraphics[width=0.65\columnwidth]{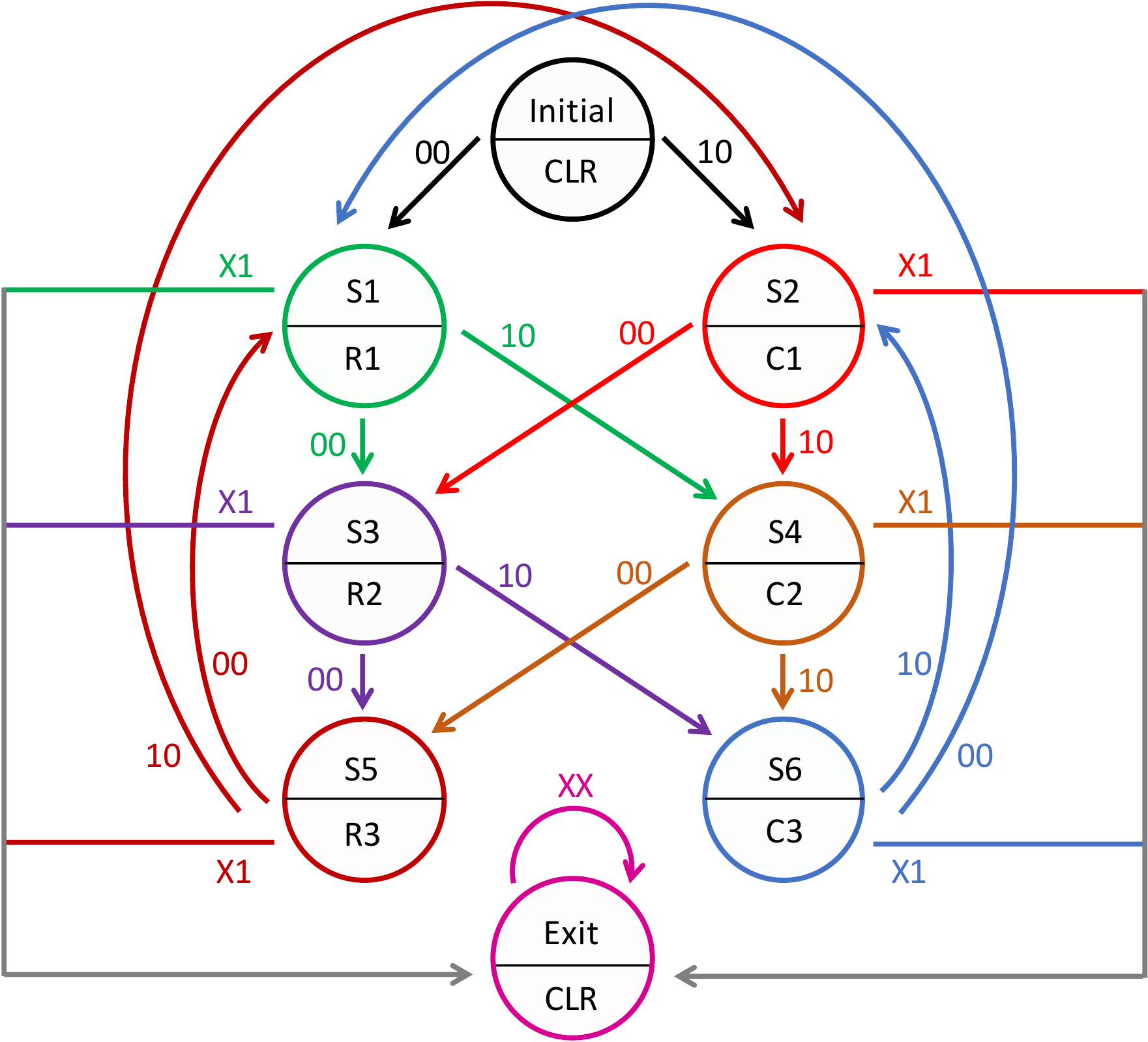}
\caption{State diagram of the finite state machine.}
\label{fig:FSM}
\end{figure}

Algorithm~\ref{alg:2} shows at a high level the pattern detection procedure for a length-$p$ pattern. Fig.~\ref{fig:pattern_detector} is a block diagram of the full pattern detector. In addition to the input X from the PISO register, another input, D, is needed so that the pattern detector can determine the end of the sequence. The pattern detector consists of a finite state machine, three pointer blocks, a comparator, and a global maximum register. The finite state machine, which takes inputs X and D, outputs the signals C and R needed for each pointer block. A clear output (CLR) is also used to reset the counters of the pointer blocks in the initial and exit states, as will be explained later. Fig.~\ref{fig:FSM} is a state diagram of the finite state machine.
Each pointer has a block consisting of a counter, a comparator, and a maximum register. The counters are initially reset. Whenever a 1 is received from the PISO, the counter of the corresponding pointer is incremented by 1 ($C=1$), whereas if the input is 0, the corresponding maximum register is compared to the counter value and updated if needed, and the counter is then reset ($R=1$). When the sequence ends, indicated by the D input, all the counters are reset and the maximum of the three maximum registers is found and stored in the global maximum register.

\begin{algorithm}
\caption{Pattern detection phase}\label{alg:2}
\SetKwInput{KwInput}{Input}                
\SetKwInput{KwOutput}{Output}              

\KwInput{Memory array mem(m, n)}
\KwOutput{Global\_max: maximum number of consecutive pattern occurrences}

ctr = zeros(1, p)

max = zeros(1, p)

\For{$i \gets 1$  \KwTo $m$} {%

\For{$j \gets 1$  \KwTo $n$} {%
    idx = (i-1)n + j 
    
    x = (idx mod p) 
    
    \If{$mem[i,j]==1$}{
        ctr[x]++ }
    
    \Else{
        \tcc{Update the maximum}   
        max[x] = ctr[x]
    
        \tcc{Reset the counter}   
        ctr[x] = 0 }
}
}
Global\_max = maximum(max)
\end{algorithm}

As an example, let us assume that the input sequences for X and D are ``1 0 1 1 1 0 0 0 0'' and ``0 0 0 0 0 0 0 0 1'', respectively. Table~\ref{tab:pattern_detector} shows the current state starting from the initial state and each subsequent state depending on the input xD, along with the required action in that state. The last state is the exit state, which returns the global maximum.

\begin{table}[!ht]
\caption{Pattern detector actions based on the example input sequence.}
\centering
\renewcommand{\arraystretch}{1.4}
\resizebox{\columnwidth}{!}
{
\begin{tabular}{ccc}
\toprule
Current state; Input & Next state & Action in next state                                                \\ 
\midrule
Initial; 10          & S2         & Increment ctr1 (C1=1): ctr1=1                                                            \\ 
S2; 00               & S3         & Update max2: max2:=0 \newline Reset ctr2 (R2=1): ctr2=0 \\ 
S3; 10               & S6         & Increment ctr3 (C3=1): ctr3=1                                                            \\ 
S6; 10               & S2         & Increment ctr1 (C1=1): ctr1=2                                                            \\ 
S2; 10               & S4         & Increment ctr2 (C2=1): ctr2=1                                                            \\ 
S4; 00               & S5         & Update max3: max3:=0 \newline Reset ctr3 (R3=1): ctr3=0 \\ 
S5; 00               & S1         & Update max1: max1:=2 \newline Reset ctr1 (R1=0): ctr1=0 \\ 
S1; 00               & S3         & Update max2: max2:=0 \newline Reset ctr2 (R2=0): ctr2=0 \\ 
S3; 01               & Exit       & Global max = 2                                                                           \\ 
\bottomrule
\end{tabular}
}
\label{tab:pattern_detector}
\end{table}

\subsubsection{Control Unit}\label{sec:CU}
The role of the control unit is to generate the control signals required by each block of the full architecture to organize the phases and ensure the correct sequential flow of the data. For example, it generates the read, write, and reset signals in the match-index memory depending on the different memory modes to ensure the operations are correct. The control unit can be realized on software running on a coprocessor. Hence, we have not discussed its circuit implementation.

\paragraph*{Timing considerations}
When an input equal to 0 is received from the PISO register, two actions need to be done: (1) Comparing and updating the maximum register and (2) resetting the counter of the corresponding pointer. It is important to separate the comparison and reset steps in time to ensure that the comparison happens before the counter is reset. Since every pointer is activated every three cycles, we delayed the comparison by one clock cycle and the reset by two clock cycles, so that both are completed before the same pointer hits another input. When the sequence of bits is finished, the pattern detector should read four extra zeros to ensure that all the maximum registers are updated. An additional 0 is needed to move it to the exit state, when all the counters are reset and the three maximum registers are compared to determine the maximum among them. Hence, we delayed the end of the sequence signal by four clock cycles. Since the reset is delayed by two clock cycles, we used the CLR signal, which we set to 1 in the initial and exit states so that the counters are reset in those states. 

\begin{figure*}[tb]
\centering
\includegraphics[width=1.9\columnwidth]{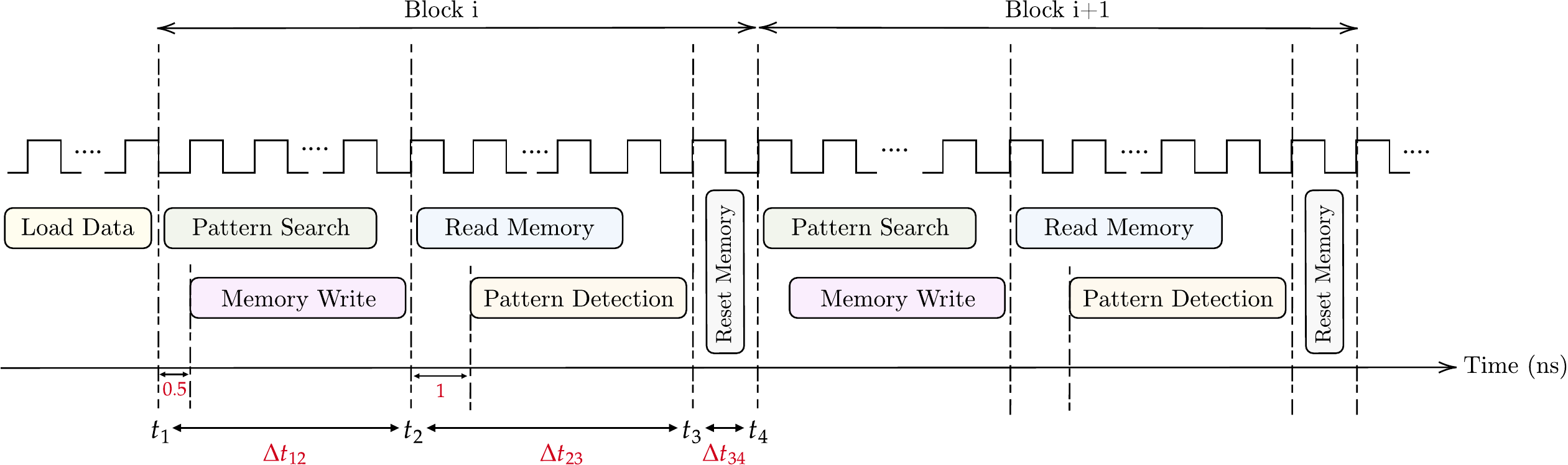}
\caption{Complete timing diagram for a full load and search cycle.}
\label{fig:timing_diagram}
\end{figure*} 

We consider that the DNA data are loaded to AM at $t_1$ seconds. The aCAM cells can be programmed row by row, so within the same row, the cells that store the same character can be programmed simultaneously. Since there are four DNA characters and two memristors in each aCAM cell, the time to load each row is $8T_w$ where $T_w$ is the memristor write time. Hence, loading the data into $M$ rows takes $8MT_w$. Note that loading the data is done only once. Then, for each block in the aCAM array, the memory write can happen in parallel with the pattern search phase, starting after a 1/2 clock cycle. After that, we read the match-index memory, and the pattern detection starts one clock cycle later. Finally, we have to reset the memory, which is done in one clock cycle. This functionality is illustrated in Fig.~\ref{fig:timing_diagram} and can be expressed by the following set of definitions:
\begin{subequations}
\begin{equation}\label{eq:1a}
t_1 = 8 M T_w
\end{equation}
\begin{equation}\label{eq:1b}
\Delta t_{12} = (N-(p-1)+0.5)T
\end{equation}
\begin{equation}\label{eq:1c}
\Delta t_{23} = 0.125\,(m\,n+5)T 
\end{equation}
\begin{equation}\label{eq:1d}
\Delta t_{34} = T
\end{equation}
\begin{equation}\label{eq:1e}
t_{s,k} =(t_1+K\,\Delta t_{14})T
\end{equation}
\end{subequations}
where $T$ is the clock period and $t_{s,k}$ is the total search time for reading $K$ blocks.

\section{Experimental Setup}\label{sec:4}
We built a simulation framework with MATLAB and HSPICE for both circuit and system simulations. In the circuit simulations, we relied on predictive technology models for 45-nm high-k/metal gate CMOS devices \cite{ptm} to study the design metrics. We set $V_{DD}=0.8$~V.

The AM has a block selector, drivers, aCAM array, and tag. We implemented the block selector using CMOS gates, as explained in Supplementary Note~1. The output of each block selector is connected to 64 drivers with three inverters each, which are connected to the rows of the aCAM array. We used an $512 \times 130$ aCAM array to store the DNA data, which we divided into eight blocks, each consisting of 64 rows. The ML of each row is connected to a driver with two inverters, followed by a flip-flop that stores the ML result.

For our memory simulations, we adopted a linear memristor model. We studied a $64 \times 128$ memory array. Accordingly the memory components were as follows.
For a write operation to a $64 \times 128$ memory array, the column selector circuit has a 7-bit counter, then a 7- to 128-bit decoder followed by 128 inverters and 128 transmission gates. To read the memory, the row selector circuit consists of a 6-bit counter, followed by a 6- to 64-bit decoder, then 64 transmission gates. To read every eight consecutive cells in parallel, we used eight 16-bit multiplexers connected to eight current sense amplifiers, whose outputs are passed to an 8-bit PISO register. The circuit connections are as shown in Fig.~\ref{fig:mem_arch}. For the column and row selector circuits, we used synchronous master--slave JK-flip-flop up counters to minimize the delays and avoid glitches. For the current sense amplifier, we used the circuit presented in \cite{csa}. We modified the $R_\text{on}$, $R_\text{off}$, and $R_\text{ref}$ parameters to meet the specifications of our circuit: $R_\text{on}=5$~k$\Omega$, $R_\text{off} = 2.5$~M$\Omega$, and $R_\text{ref}=14$~k$\Omega$.  
For the pattern detector, first, for the finite state machine, we used dynamic flip-flops to represent the states, which enhanced performance and speeded up the latching. Then, for each pointer block, we used an 8-bit D-flip-flop counter, an 8-bit comparator, and an 8-bit maximum register, as shown in Fig.~2 in Supplementary Note~2. The comparison logic, which returns the maximum value among all the maximum registers, was implemented using two 8-bit comparators, two multiplexers, and an 8-bit global maximum register, which stores the global maximum, as illustrated in Fig.~3 in Supplementary Note~2.

\section{Results and Discussion}\label{sec:5}
In this section, we evaluate the performance of our circuit in terms of key metrics, namely latency, energy, and area.
First, for the delay, by applying the time equations in the timing considerations under Section~\ref{sec:CU} and based on Fig.~\ref{fig:timing_diagram}, assuming $T_w = 1$~cycle, the amount of overhead needed to load the data into the aCAM array is $8 M T = 4.096 ~\mu$s. The memory write and pattern search time is 128.5~ns. The memory read and pattern detection time is 1024.625~ns, and the memory reset time is 1~ns. As such, in total, ignoring the time overhead, the pattern matching task for one aCAM block takes around $1.15~\mu $s. Therefore, our design significantly reduces the time required for DNA pattern matching.

Energy values were obtained from simulations, except for writing the memory, since we adopted a linear resistor model, so we assumed that the memristor SET energy is 1~pJ. We calculated that the full write energy of the 1T1R memory array is 1.228~nJ. The reset energy is the same as the write energy, since all the SET cells need to be reset. The memory read energy is 0.82~nJ, the pattern search energy is 1.1769~nJ, and the pattern detection energy is 0.7709~nJ. Thus, the total search energy per character is 0.61~pJ.

\begin{figure} [!ht]
\centering
\subfigure[]{\includegraphics[width=0.445\columnwidth]{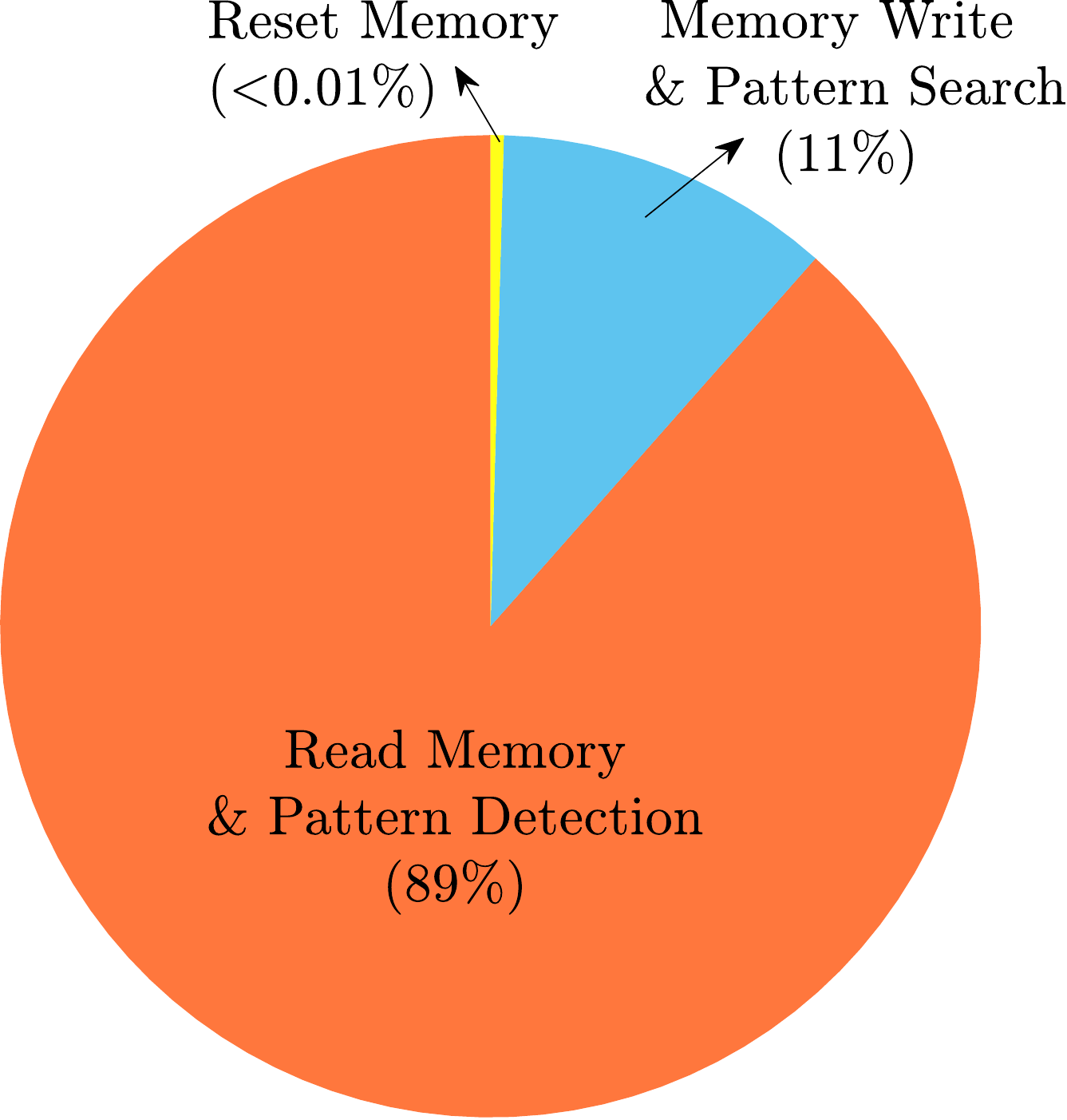}}
\subfigure[]{\includegraphics[width=0.4\columnwidth]{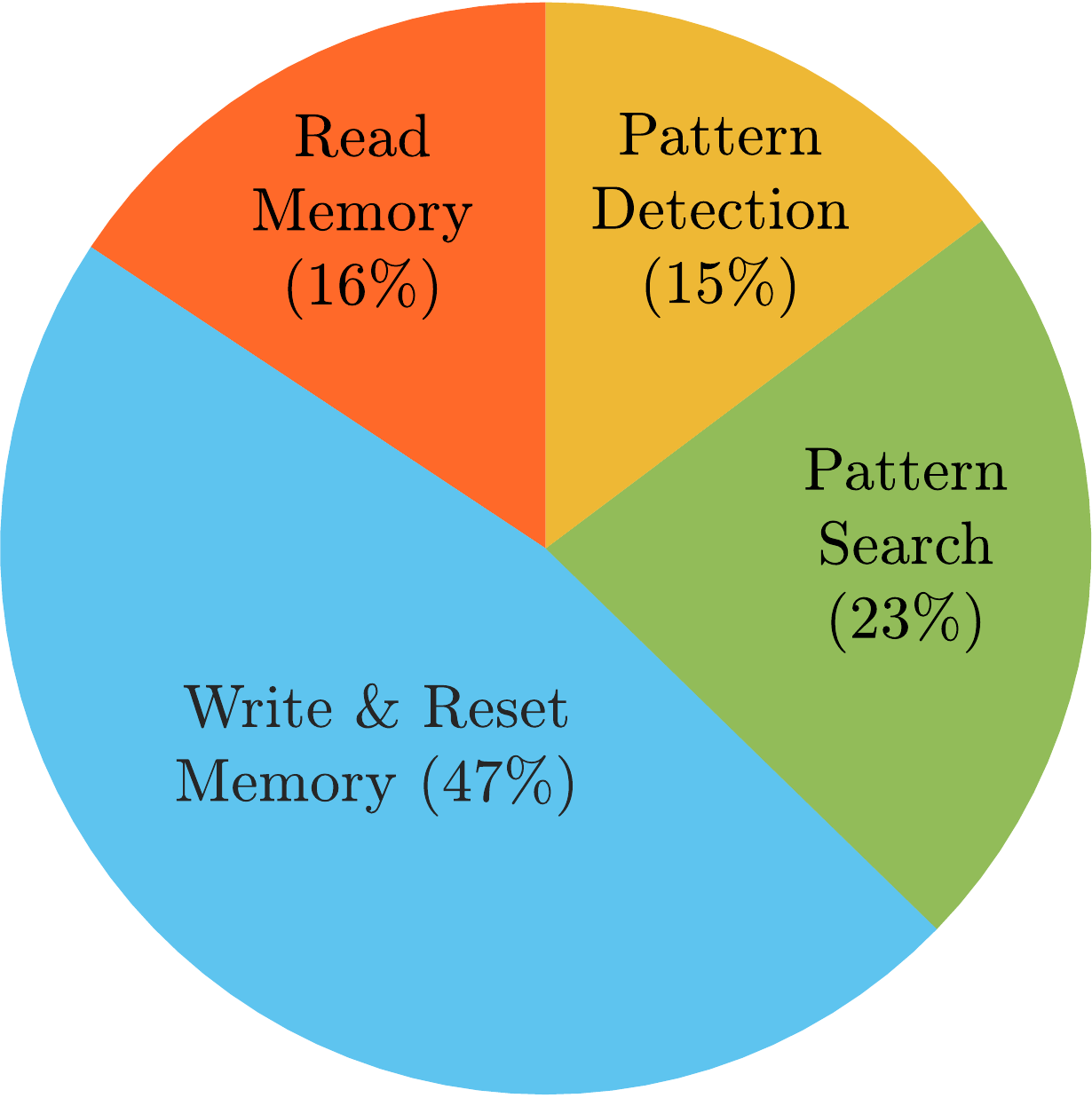}}
\caption{Simulation results: (a) latency breakdown and (b) energy breakdown. }
\label{fig:piechart}
\end{figure}

Fig.~\ref{fig:piechart} shows the distributions of both time and energy among the different operations. Note that the read memory and pattern detection modes as well as memory write and pattern search are combined in the latency figure since they operate in parallel.

The active area estimates are based on the calculations presented in Supplementary Note~4. Accordingly, the AM area was calculated to be 47.78~mm$^2$, the match-index memory area as 0.155~mm$^2$, and the pattern detector area as 0.0926~mm$^2$.

To determine the effect of increasing the pattern length on run time and energy, we consider the example of a DNA sequence of 1 million characters.
For a pattern of length 3, there are two replicated cells in each row, so 16 aCAM arrays are needed to store the DNA data, each consisting of eight blocks. Assuming that we have one shared memory for all the blocks of the aCAM arrays, for $p=3$, ignoring the time overhead in Equation~\eqref{eq:1e}, the total run time for the pattern matching task is 147.7$~\mu$s. On the other hand, for a pattern of length $p=5$, we still need 16 aCAM arrays. However, in this case, there are four replicated cells in each row, hence $130-4=126$ cycles instead of 128. The total run time is 144.4$~\mu$s. Increasing the pattern length to 10, and following the same logic, the total run time is 148.393$~\mu$s.
Moreover, we estimated the energy per cycle for each block in our design based on our simulations. For each pattern length, we multiplied the energy values per cycle by the corresponding number of cycles. Accordingly, for $p=3$, the total energy is 5.2~nJ. For $p=5$, it decreases slightly to 5.09~nJ, and for $p=10$, it is 4.9~nJ.


\section{Conclusion}\label{sec:6}
In this paper, we propose a hardware accelerator for DNA pattern matching. The proposed design stores the DNA characters in AM and accelerates a hardware-friendly algorithm in a highly parallelized way. The algorithm counts the maximum number of consecutive repeats of a specific pattern, which can be used to detect the presence of possible trinucleotide repeat-expansion genetic disorders. The proposed design is energy-efficient, and it showed a remarkable improvement in terms of time cost compared to software implementations of DNA pattern matching. The fabrication of this hardware accelerator is left for future work. Moreover, we are planning to test the proposed design on practical datasets containing human genome sequences.



\end{document}